# Nonadiabatic electron dynamics effects on high-harmonic generation spectrum of $H_2^+$: minima and oscillatory pattern

S. Taghipour, M. Vafaee*

*Department of Chemistry, Tarbiat Modares University, PO Box 14115-175, Tehran, Iran*

We numerically solved the full-dimensional electronic time-dependent Schrödinger equation for $H_2^+$ with Born-Oppenheimer approximation under different sin$^2$-shaped and trapezoidal laser pulses at some different wavelengths, with $I = 1 \times 10^{13}$, $3 \times 10^{13}$ and $6 \times 10^{13}$ Wcm$^{-2}$ intensity at 4.73 a.u. and 7.0 a.u. internuclear distances. Some structures such as complexity, minima, and oscillatory patterns appeared in the high-order harmonic generation (HHG) spectra are investigated in this work by considering the electron localization, electron nonadiabatic dynamics, spatially asymmetric of the HHG, and the Rabi frequency of the population of the ground and excited electronic states to better understand the origins of these structures in the HHG spectrum. We will clear that the origin of complicated patterns of the HHG spectra in sin$^2$-shaped laser pulse is due to that the most portion of the HHG emission occurs at the falling part of the laser pulse. We explore that the oscillatory pattern in the HHG spectra originate from an oscillatory pattern in the $S_g$ and $S_u$ spectra and these oscillatory patterns in turn are due to the nonadiabatic electronic behavior appeared as the slow oscillation pattern in the ground and first excited electronic states populations. Also, our result shows that the minima of the HHG are related to the oscillatory patterns in $S_g$ and $S_u$ spectra.

## I. INTRODUCTION

Observation of the electron dynamics requires ultrashort lasers around the attosecond time scale (~10$^{-18}$ s) [1, 2]. Electron processes can be evaluated and modified by attaining this time resolution. One of the ways to achieve such ultra-short lasers is high-order harmonic generation (HHG). When an atom or molecule is under the intense laser pulse, photons with multiple of the initial fundamental frequency of the laser pulse can be generated. Such a phenomenon is called high-order harmonic generation [3]. Nowadays, HHG is the traditional way to produce spatially and temporally coherent extreme ultraviolet radiation (XUV) light, as well as light source in the attosecond regime [4]. Recently, to access such time scale and also to apply in some fields such as control of electron wavepackets [5], molecular tomography [6], and easier access to X-rays [7], HHG has been gained much attention in the science of laser-matter interaction [4].

A simple model that allows to understand the basic features of HHG can be given by the following three-step model [3]. In the first step, the laser field distorts the Coulomb potential, so that the electron can tunnel out of the Coulomb attraction and can be free. In the second step, the ionized electron is born with zero velocity and accelerated in the laser field. When the direction of the laser field reverse, the free electron may recombine with the ion depending on the phase of the field at its birth time, and the energy is released as HHG emission (third step). At recollision, a single photon is emitted with energy $E = I_P + k$ that $I_P$ is the ionization potential and $k$ is the kinetic energy upon recollision. This process can repeat for each half of the laser cycle, which results in an attosecond pulse train for a laser pulse with several optical cycles [4, 8]. The HHG spectrum of molecules usually has specific structures such as non-odd harmonics with complexities in some part of the spectrum, the sudden decrease of intensity in some harmonic orders as minima, the blueshift, redshift, and intramolecular interference.

The blueshift and redshift of harmonics occur in rising and falling parts of the laser pulse, respectively. This is one of reasons to appear complexities in the HHG spectrum. Bian et al showed that for laser pulses sin$^2$ and Gaussian envelopes consists of HHG emission mainly in the rising or falling part, respectively the blueshift or redshift of harmonic orders occurs relative to odd harmonic orders [9]. The imbalance of the HHG generation on the falling and rising parts of the laser field leads to a blueshift or redshift in HHG spectrum.

Ahmadi et al. recently reported a trapezoidal laser pulse with two-cycle falling part that led to a complexity of the HHG spectrum and a non-odd order harmonics were observed [10]. They showed that a nonadiabatic response of the molecule to the rapidly changing laser field and a spatially asymmetric emission along the polarization direction mainly leads to the modulation of the HHG spectrum [10].

In this work, we numerically solve the full-dimensional electron dynamics of $H_2^+$ under sin$^2$-shaped and trapezoidal laser pulses at some different wavelengths, with $I = 1 \times 10^{13}$, $3 \times 10^{13}$ and $6 \times 10^{13}$ Wcm$^{-2}$ intensity at 7.0 a.u. and 4.73 a.u. nuclear distances and derived the HHG spectra. Our focus is to uncover the origin of the complexities, oscillatory pattern and minima in these HHG spectra. In order to understand and identify the underlying physics behind these structures in depth, we perform an analysis of nonadiabatic electronic behavior and investigate this effect on the time-dependent population of the electronic states. Recently particular attentions have been paid to study nonadiabaticity in electron dynamics of the HHG spectra [10-13]. Miller et al [12] related these complexities of the HHG spectra to the change of the electron localized pattern in the laser field. They reported that recombination events are occurring near the center of the driving laser. They stated at the nearly constant amplitude of the laser pulse, the electron dynamics is highly nonadiabatic and several localizations occurs per half-cycle of the laser field. They proposed that, at the nearly constant amplitude of the laser pulse, if the electron repeats its non-adiabatic behavior in each period, the resulted HHG spectrum would have been expected odd orders [12].

Lein et al. stated that maxima and minima can be considered as the result of interference between two radiating point sources located at the nuclei. [14]. For describing the structure of minimum interference, theoretical [14-16] and experimental [6,17-25] studies have been performed. Han et al. studied the role of the internuclear distance on the interference minimum on the HHG spectrum of $H_2^+$ [26]. They reported that when the internuclear distance is increased, contribution of recombination into the first excited state plays important role and has not been neglected and the orbital interference term also needs to be taken into account that leads to the failure of the two-center interference model.

Recently, the relationship between minimum and the transient of the electron localization on the HHG spectrum have discussed by Miller et al [11, 12]. They argued that the minimum of HHG spectrum is related to the phase difference between the electron emission and the remained wave packet at the time of recombination [11]. They stated that non-adiabatic dynamics is closely related to the time-dependent phase of the electron wave packet $\psi(\rho, z; t)$ around each nucleus [11]. They explained that at high wavelengths (such as 1400 and 1800 nm), minimum of HHG spectrum is related to the wave-packet phase ($\alpha$) at the time of ionization and is not correlated with the electrons wave packet phase in the recombination time. They also expressed that when $\alpha$ is zero at the ionization time, the suppression would be occurred in the ionization and consequently in the generation of the HHG spectrum which leads to appear a minimum in the HHG spectrum. In contrast, when $\alpha$ is not zero at the ionization time, there is no suppression on the HHG spectrum [11].

In this work we seek underlying physics behind the harmonic emission in $H_2^+$ under relatively nonadiabatic electron dynamics. We will analysis the effect of resonance between ground and excited electronic states on the HHG spectra and investigate the role of these two lowest electronic at different internuclear distances, wavelength, intensity, and envelope of laser pulse. For study of the origin of the oscillatory patterns on the HHG spectrum, we investigate the HHG spectrum due to the ground and first excited electronic states. The rest of the paper is organized as follows: Section 2 describes the framework for the numerical methods used to solve the time-dependent Schrödinger equation (TDSE) for $H_2^+$ under the laser pulses. In Section 3, simulation results are presented and discussed. Finally, Section 4 presents the conclusions. We use atomic units throughout the article unless stated otherwise.

## II. NUMERICAL RESULT

Time-depended Schrödinger equation (TDSE) for a fixed-nuclei model of $H_2^+$ exposed to an external linearly polarized electric field can be expressed (in atomic units; $e = \hbar = m = 1$) as [27, 28]

$$i\frac{\partial \psi(z,\rho;\,t,R)}{\partial t} = H(z,\rho;\,t,R)\,\psi(z,\rho;\,t,R), \quad (1)$$

with electron cylindrical coordinate $(z, \rho)$ which are measured with respect to the center of mass of the two nuclei (after a separation of the center-of-mass motion and ignoring molecular vibration and rotation). $H$ is the total electronic Hamiltonian for $H_2^+$

$$H(z,\rho;\,t,R) = -\frac{2m_n + m_e}{4m_n m_e}\left[\frac{\partial^2}{\partial \rho^2} + \frac{1}{\rho}\frac{\partial}{\partial \rho} + \frac{\partial^2}{\partial z^2}\right] + V(z,\rho;\,t,R), \quad (2)$$

$m_e$ and $m_n$ are the masses of electron and a single nuclei, respectively, with

$$V(z,\rho;\,t,R) = -\frac{1}{\sqrt{\left(z+\frac{R}{2}\right)^2 + \rho^2}} - \frac{1}{\sqrt{\left(z-\frac{R}{2}\right)^2 + \rho^2}}$$
$$+\frac{1}{R} + \left(\frac{2m_n + 2m_e}{2m_n + m_e}\right) zE(t). \quad (3)$$

In this equations, the electric field is defined $E(t) = E_0 cos(\omega t) sin^2(\pi t/N\tau) cos(\pi t)$ and $\omega$ is angular frequency. In this work we used of the $sin^2$-shaped laser pulse with 20 o.c. (units of "o.c." mean the optical cycle of the pulse); is shown in Fig. 1 and a trapezoidal pulse envelope with of time duration 25 o.c. with 10 cycles ramp on, 5 cycles constant, and 10 cycles ramp of that shown in Fig. 3. The TDSE is solved using unitary split-operator methods [29, 30] which the detailed numerical procedures can be found in Refs [31-33]. The finest grid size values in our numerical integration are 0.13 and 0.2, respectively for $z$ and $\rho$. The size of the simulation box is chosen as $z_{max} = 157$ and $\rho_{max} = 124$.

The time-dependent wave function was used to obtain the power spectrum of the HHG radiation by calculating the square of the windowed Fourier transform of dipole acceleration $a_z(t)$ in the electric field direction $(z)$ as

$$S(\omega) = \left|\frac{1}{\sqrt{2\pi}}\int_0^T \langle \psi(z,\,\rho;\,t,R)|a_z(t)|\psi(z,\,\rho;\,t,R)\rangle_{z,\rho} \times H(t)exp[-i\omega t]dt\right|^2, \quad (4)$$

where

$$H(t) = 1/2[1 - cos(2\pi t/T)], \quad (5)$$

$H(t)$ is the Hanning filter and $T$ is the total pulse duration. We use the Hanning filter to reduce the effect of unphysical features

on the HHG spectrum that last after turn-off of the laser pulse. The spatial distributions of corresponding HHG spectra as a function of the electronic coordinate $z$, $S(z, \omega)$ is given by

$$S(z,\omega) = \left| \frac{1}{\sqrt{2\pi}} \int_0^T \langle \psi(z, \rho; t, R)|a_z(t)|\psi(z, \rho; t, R) \rangle_\rho \times H(t) exp[-i\omega t]dt \right|^2. \quad (6)$$

To calculate contributions of different electronic states to total HHG spectrum, $\psi(z, \rho; t, R)$ can be separate into the following components [34]:

$$\psi(z, \rho; t, R) = c_g(t)\psi_g(z, \rho; t, R) + c_u(t)\psi_u(z, \rho; t, R) + \psi_{res}(z, \rho; t, R), \quad (7)$$

where $\psi_g(z, \rho; t, R)$ and $\psi_u(z, \rho; t, R)$ refers to the wavefunctions of the ground and first excited states, respectively, corresponding to the $1s\sigma_g$ and $2p\sigma_u$ states. $\psi_{res}(z, \rho; t)$ is related to the residual part of total wavefuntion $\psi(z, \rho; t, R)$ containing other excited and continuum states. With substitute Eq. 7 to Eq. 4 and retain the dominant terms, we can write

$$S_{tot}(\omega) \approx S_{gu}(\omega) + 2[A_g^*(\omega)A_u(\omega)] \quad (8)$$

$$S_{gu}(\omega) = S_g(\omega) + S_u(\omega)$$

where $S_g(\omega) = |A_g|^2$, $S_u(\omega) = |A_u|^2$ and

$$A_g(\omega) = \int 2Re \langle c_g(t)\psi_g(t)|a_z(t)|\psi_{res}(t) \rangle e^{i\omega t}dt, \quad (9)$$

$$A_u(\omega) = \int 2Re \langle c_u(t)\psi_u(t)|a_z(t)|\psi_{res}(t) \rangle e^{i\omega t}dt. \quad (10)$$

In these relations, $S_g(\omega)$ and $S_u(\omega)$ refers to recombination to the $1\sigma_g$ and $2p\sigma_u$ respectively, and the term $2[A_g^*(\omega)A_u(\omega)]$ corresponds to the electronic interference term of these two localized electronic states. If $S(\omega) \approx S_{gu}(\omega)$ then the HHG spectrum includes recombination to the ground state, $S_g(\omega)$, and first excited state, $S_u(\omega)$.

In this work, the third and fourth excited electronic states are not considerably populated during the interaction and therefore the corresponding terms $S_3(\omega)$ and $S_4(\omega)$ is negligible. To study the time profile of harmonics generated, an inverse the Fourier transform over a selected range of frequencies is obtained by Morlet-wavelet transform of dipole acceleration $a_z(t)$ via [35, 36]

$$w(\omega, t) = \sqrt{\frac{\omega}{\sigma\pi^{\frac{1}{2}}}} \times$$

$$\int_{-\infty}^{+\infty} a_z(t')exp[-i\omega(t'-t)] \exp\left[-\frac{\omega^2(t'-t)^2}{2\sigma^2}\right]dt'. \quad (11)$$

We use $\sigma = 2\pi$ in this work.

## III. RESULTS AND DISCUSSION

The high-order harmonic spectrum for the $H_2^+$ system, in $R = 7.0$ a.u. internuclear distance, under a 20 cycles $\sin^2$-shaped laser field at 1400 nm wavelength and $6\times10^{13}$ Wcm$^{-2}$ intensity (shown in Fig. 1) has an interesting and special structure in the plateau which can be seen in Fig. 2. As seen in Fig. 2, there are an oscillatory behavior and a complexity between the 48th to the 60th harmonic orders on the HHG spectrum. To investigate about the origin of this complexity of the HHG spectrum, we have examined the HHG spectrum emitted by a trapezoidal pulse of 25 optical-cycle duration (10 cycles ramp on, 5 cycles constant, and 10 cycles ramp shown in Fig. 3) at 1400 nm wavelength and $6\times10^{13}$ Wcm$^{-2}$ intensity. Figure 4 shows the HHG spectrum of the $H_2^+$ system with $R = 7.0$ a.u. internuclear distance for this trapezoidal laser field. It can be seen that the complexity (between 48-60 harmonic orders) is disappeared but there is still the oscillatory behavior.

Ahmadi et al shows that the origin of complicated patterns of HHG spectra is due to that the most portion of the HHG emission occurs at the falling part of the laser pulse [10]. In order to understand the role of the envelope shape of these laser pulses in the complexity behavior of HHG spectrum, we plotted in Fig. 5 the corresponding Morlet wavelet time profile of the HHG spectra (Eq. 11) of Fig. 2 and Fig. 4. For Fig. 5(a), corresponding to the $\sin^2$ laser pulses in Fig. 2, the harmonics are generated mostly in falling part (the second ten-cycle) and according to Ahmadi's statement [10], we expect its HHG becomes complicated. In contrast, for the Fig. 5(b) corresponding to the trapezoidal laser pulses in Fig. 4, one can see the significant of the harmonic orders generated in the constant amplitude part of the envelope laser field (10 to 15 cycles). Therefore, we can expect that the complexities of Fig. 2 disappear in Fig. 4.

Looking to find out a correlation between the complexities and non-odd harmonics in Fig. 2 and the absent non-odd harmonics in Fig. 4, we also performed the calculations of the spatial distribution of HHG spectra as a function of the electronic coordinate $z$ ($S(z, \omega)$) in Eq. 6, as shown in Fig. 6. Fig. 6(a) shows $S(z, \omega)$ of the complexity part of HHG spectrum in Fig. 2 for the $\sin^2$ field and Fig. 6(b) shows the corresponding part of HHG spectrum in Fig. 4 for the trapezoidal laser field. From Fig. 6(a), we can observe that the symmetry of the HHG emission is broken for $z < 0$ and $z > 0$ regions. But in Fig. 6(b), the HHG for these two regions is nearly symmetric for both odd and even harmonic orders.

For establishing a connection between the complexity patterns in the HHG spectra and the laser-driven transient localization of the electron wavepacket, we plot in Fig. 7 the electron localization for the $\sin^2$-shaped laser field (Fig. 1) and the trapezoidal laser field (Fig. 3). For the $\sin^2$-shaped laser field, it shows an adiabatic behavior from 1 to 4 and 16 to 20 cycles and a non-adiabatic behavior from about 4 to 16 cycles. Here, nonadiabatic refers to the driven electron motion with unlike direction respect to the oscillations of the electric field during a laser cycle. For the trapezoidal field, non-adiabatic behavior is from about 4th period until the 20th period, and adiabatic behavior is for others periods (1 to 4 and 21 to 25 cycles). In the right panels of Fig. 7, during some cycles it can see the electron localization loses its periodicity. In the Fig. 7(b), a repetitive pattern of the non-adiabatic behavior of the electron localization can be seen form about 8th until the 12th periods (near the center of the driving laser), but after this time, this repetition pattern is deformed. Miller et al. proposed that the complexities of the HHG spectrum is due to disappearance of this repetitive pattern [12]. Miller et al stated that the recombination events occurring near the center of the driving laser. At this moment, the electron dynamics are highly nonadiabatic. They expressed that if the electron nonadiabaticity remains same from one cycle to the next cycle, odd harmonics are generated. But, if the periodicity of the nonadiabatic electron dynamics changes to an aperiodic nonadiabatic dynamics, even harmonics is appeared. In the Fig. 7(d) for the trapezoidal field, the electron localization has non-adiabatic behavior and repetitive trend form about 8th until the 17th periods. After this time, the repeating pattern is deformed similar to sin2-shaped laser field but the HHG spectrum does not show complexity. As we showed before in Fig. 5, we can relate the complexity pattern of the HHG spectrum to the significant emission of HHG in the falling part of the laser pulse. Also, recently the origin of the HHG spectrum complexity in Fig. 2 is proposed related to increased nonadiabatic character of the electronic intramolecular motion that is coincide with a triple-peak structure in the localization distribution per half-cycle throughout the central cycles of the laser field as shown in Fig. 7(a,b) [12]. Considering Fig. 7(c,d) for the trapezoidal laser field, it can be seen that the despite the appearance of the nonadiabatic character of the electronic motion and also the appearance of the triple-peak structure in the localization per half-cycle at the constant amplitude of the trapezoidal laser pulse, but it was not observed the complexity on the HHG spectrum in Fig. 4.

As mentioned, there is an oscillatory behavior on the HHG spectrum of Fig. 2 and 4. Figure 8 shows the comparison between $S(\omega)$ and $S_{gu}(\omega)$ (Fig.8(a)), and between $S_g$ and $S_u$ (Fig.8(b)), and $S_g$ (Fig.8(c)) and $S_u$ (Fig.8(d)) at $R$=7.0 a.u. internuclear distance for the sin$^2$ laser field (Fig. 2) at 1400 nm wavelength with $I = 6 \times 10^{13}$ Wcm$^{-2}$ intensity. Figure 8(b) shows the oscillatory pattern on the HHG spectrum with the blue and red double-arrows that related to $(S_g)$ and $(S_u)$, respectively. Figure 8(a) shows that $S_{gu}(\omega)$ (the sum of $S_g$ and $S_u$) is equal to $S(\omega)$. Therefore, the total HHG spectrum in Fig. 8 is mostly relate to the ground and the first excited electronic states, and the role of the other excited electronics states in the spectrum are negligible. It can be seen that the oscillation of $S(\omega)$ in Fig. 8(a) is related to the oscillation of the spectra of $S_g$ and $S_u$. For different parts of $S(\omega)$, one of these two $S_g$ and $S_u$ overcomes as shown in Fig. 8(b) with the blue arrow for $S_g$ and the red arrows for $S_u$. These oscillations of $S_g$ and $S_u$ are also seen for the trapezoidal envelop shape shown in Fig. 3 with 1800 nm wavelength and $I = 6 \times 10^{13}$ Wcm$^{-2}$ intensity. From Fig. 9, one can see that for different parts of $S(\omega)$, one of these two $S_g$ and $S_u$ overcomes and the oscillatory pattern in $S(\omega)$ due to the oscillation of the spectra of $S_g$ and $S_u$. Therefore, these oscillations of the of $S_g$ and $S_u$ are responsible for the oscillatory patterns observed at the HHG spectra in Fig. 2 and Fig. 4. To better representation of the oscillatory patterns and contribution of the ground and excited electronic states on the HHG spectra, $S_g - S_u$, the difference between $S_g$ and $S_u$, is shown in Fig. 10 (a,b) that corresponding to Fig. 2 and Fig. 4, respectively. It can be seen that $S_g - S_u$ has clearly an oscillatory behavior. For example, Fig. 10 (a) shows that $S_u$ for the harmonic orders between 1-13, 30-40, and 70-80 is dominated. In Figs. 11 and 12, we examine the effect of internuclear distance and wavelength on the oscillatory behavior of the HHG spectra under the sin$^2$ and trapezoidal laser pulses. Figure 11 shows the HHG spectra for the $H_2^+$ system, in $R$=7.0 a.u. internuclear distance at 1800 nm wavelength. Comparison of Fig. 8, Fig. 9 and Fig. 11 shows that by increasing wavelength, the cutoff position is increased. However, the similar oscillatory pattern is seen on the HHG spectrum in Fig. 11. In contrast with the $R = 7.0$ a.u. where the ground and excited electron states nearly degenerate, these two states are not degenerate in the internuclear distances of 4.73 a.u. Figure 12 represent the total HHG spectrum (left panels), the HHG of the ground and the first excited states (right panels) for the sin$^2$ and trapezoidal laser pulses at 1400 and 1800 nm wavelengths at 4.7 a.u. internuclear distances. It can be observed in all of them the oscillatory patterns of $S_g$ and $S_u$ spectrum are similar to Fig. 8 and 10 for $R$=7.0 a.u. As a result, the oscillatory pattern of $S(\omega)$ spectrum are due to the contribution of the oscillatory patterns of the ground and first excited electronic states.

To proceed, we now investigate the relation between the oscillatory pattern in HHG spectra and the time-dependent population. Figures 13 and 14 show the time-dependent population of, respectively, the ground and first excited electronic states of the $H_2^+$ system at $R = 7.0$ a. u., under different duration (5, 10, 15, 20, 25, and 30 cycles) of sin$^2$-shaped laser pulses with the wavelength 1400 nm and $I = 6 \times 10^{13}$ intensity shown in Fig. 15. When a system is interacting with a laser pulse, the population of the initial ground electronic state is transmitted to other electronic states or release as electron wavepackets into the continuum states. For these laser pulses shown in Fig. 15, mainly the first two electronic states (ground and first excited electronic states) of the $H_2^+$ at $R = 7.0$ a. u. have population. The oscillation of the population in Figs. 13 and 14 results from the population transmission between the ground and first electronic excited states under the interaction with the laser pulse. The population starts to exchange periodically between two ground and first excited electronic states by turning on the laser pulse. As the number of cycles of the laser pulses and therefore amplitude of laser pulses increases and at near the center of the laser envelopes, the oscillation rate of the populations between the two first electronic states increases. Figures 13 and 14 show that for the all laser pulses, independent of the number of the field cycles, the population transmissions have two type oscillations; the Rabi oscillation due to coupling of the ground and first excited electronic states shown with blue lines and the slow oscillation shown with red lines. The slow oscillation is related to the variation of the Rabi frequency that in turn is due to the intensity variation during the rising and falling edge of the laser pulse. Our results show that the Rabi oscillation and slow oscillations are also appeared for the trapezoidal laser field, which we do not show here. In Figs. 13 and 14, by transition of the population to an electronic state, the probability of ionization, recombination, and therefore HHG emission increase for this state. Since the total HHG spectrum is mainly a contribution of the ground and excited states, the HHG spectrum is influenced by the smooth oscillation (red line) of the population of these two states shown in Figs. 13 and 14. This smooth oscillation of the transition of the population between the two electronic states effects both ionization/recombination rate from/to these two ground and excited electronic states which in turn affects the HHG spectrum. The effect of this smooth oscillation of the Rabi frequency appears in the oscillatory patterns of the HHG spectrum in Figs. 8, 9, 11 and 12.

To better understand the contribution of the ground and excited electronic states on the HHG spectra, the Morlet-wavelet Fourier transform of the HHG spectra of Fig. 8(b) is depicted in Fig. 16. Figures 16(a) and 16(b) are related to the HHG of the ground and the first excited electronic states between 8th and 12th field cycle. According to Fig. 13(d) and 14(d) can be seen the ground state has more population in duration of 8-12 optical cycles of the laser cycles. Figures 16(a) and 16(b) show that the contribution of HHG process from the ground state in this duration of the laser field (8-12 o.c.) is more than that from the excited state. Also, according to Fig. 13(d) and 14(d) can be seen the population of the excited state ($14 < t < 16$ o.c.) is averagely more than that the ground state. As a result, figures 16(c) and 16(d) show that the contribution of the excited state in this duration of the laser field (14-16 cycles) is more than that the ground state. Therefore, these periodic oscillations on the $S_g$ and $S_u$ lead to the oscillatory pattern of $S(\omega)$.

For sin$^2$ laser pulse for 7.0 a.u. internuclear distance at 1800 nm wavelength with $I=6\times10^{13}$ Wcm$^{-2}$ intensity, Miller et al reported a minimum as a signature of the transient localization of the electron upon alternating nuclear centers and as representative of the dynamics that occurs exclusively at the time of ionization (see Fig. 11(a)) [11]. Similar such minimum can been see in Figs. 2, 4 and 11(c). In fact, there is not only one minimum in the HHG spectra but also we can see that there is a set of minima in the HHG spectra of Figs. 2, 4, and 11. From these figures, we can relate these minima to the oscillatory pattern of the $S_g$ and $S_u$ spectra and subsequently of the $S(\omega)$ spectrum.

The Rabi frequency is the frequency of fluctuation in the populations of the two-level involved in the transition and depends on the laser intensity and wavelength [37, 38]. For example, in Figs. 13 and 14, when the intensity is increased during the rising edge of the laser pulse, the Rabi frequency is increased. The effect of the magnitude of the intensity of the laser field on the patterns of the electron localization, the ground electronic state population, and the Rabi oscillations are represented in Fig 17. This figure shows the electron wavepacket localization (a, c, e and g) and the corresponding the time-dependent population of the ground electronic state (b, d, f and h respectively) of the H$_2^+$ system at $R = 7.0$ a. u. internuclear distance under sin$^2$-shaped laser pulses and trapezoidal pulse with the wavelength 1400 nm for different $I = 1 \times 10^{13}$ and $I = 3 \times 10^{13}$ Wcm$^{-2}$ intensities. From the electron localization diagrams, it can be seen the electron motion remains relatively adiabatic throughout the whole driving laser pulses with $I = 1 \times 10^{13}$ Wcm$^{-2}$ intensity as shown in Fig. 17(a) and (e) for sin$^2$-shaped and trapezoidal laser pulses, respectively. By increasing the driving field intensity ($3 \times 10^{13}$ Wcm$^{-2}$), the adiabatic motion of the electron wavepacket changes to a nonadiabatic motion (Fig. 17(a,c) for the sin$^2$ and Fig. 17(e,g) for the trapezoidal laser pulses). We observe double peaks in the nonadiabatic behavior of Fig. 17(c, g) and changes in the related Rabi frequency of the electronic state population in Fig. 17(d, h) as a slow oscillation marked with red line. It can be seen in Fig. 17 that by increasing the driving field intensity, the rate of Rabi oscillations is increased. Another point in Fig. 17 is that when the laser field amplitude is constant for example between 10-15 optical cycles in the trapezoidal laser pulse, we do not see any change in the pattern of Rabi oscillation. In other words, when laser amplitude is constant, the pattern of the Rabi oscillation in the time-dependent population of the ground electronic state does not change.

We can also investigate the effect of wavelength of the laser pulse on the time-dependent behavior of the ground and excited electronic states populations. Figure 18 shows the time-dependent population of the ground electronic state of the H$_2^+$ at $R = 7.0$ a. u. internuclear distance under sin$^2$-shaped laser pulses at different wavelengths (800, 900, 1000, 1100, 1200, and 1300 nm) and $I = 6 \times 10^{13}$ Wcm$^{-2}$ intensity. We can see, as the wavelength increases, the nonadiabatic motion appears sooner and the Rabi frequency is increased. This effect also appears for the trapezoidal laser pulse as shown in Fig 19 for two wavelengths 1400 and 1800 nm.

Figure 20 compares the electron localization along the $z$ axis (black line) with the population of the ground and excited electronic states (blue line) for the 20 cycle sin$^2$ laser pulses at 1400 nm wavelength and $I = 6 \times 10^{13}$ Wcm$^{-2}$ intensity (shown in Fig. 1). Regarding the electron localization, it can be seen that when an electron is alternated localization from one nucleus (proton) to another nucleus (marked with the red dashed arrow as an example), subsequently a wide valley occurs in the population of the ground electronic state (marked with the red arrow). For any triplet in the electron localization curve on each nucleus (marked with the black dashed arrow as an example), we can relate a triplet peak in the population (marked with the black arrow). Similar scenario is also observed for the first excited electronic state (shown in Fig. 20(b)); for any alternated localization from one nucleus to another nucleus, subsequently a wide peak occurs in the population of the ground electronic state and for any triplet in the electron localization curve, we can relate a triple valley in the population. Figure 20(c) shows that at the onset of nonadiabatic electronic behavior (at about 4.1 optical cycle), the singlet peak is replaced by doublet in the electron localization curve and we can relate these doublets in the electron localization curve to doublet valley in the population of the ground electronic state. We can conclude that the slow variation in the Raby frequency appeared as slow oscillations pattern in the time-dependent population of the ground and first excited electronic states are due to nonadiabatic electron behavior (see Fig. 17 and 20). Therefore, it can be said that the mentioned minimum in the reports is related to the oscillations in $S_g$ and $S_u$, and in turn these oscillations are due to slow oscillation patterns in the time-dependent population of the ground and first excited electronic states and consequently the nonadiabatic electron behavior.

## IV. CONCLUSIONS

In this work, we solved numerically the full-dimensional electronic time-dependent Schrödinger equation for $H_2^+$ with Born-Oppenheimer approximation under different $sin^2$-shaped and trapezoidal laser pulses at some different wavelengths, with $I = 1 \times 10^{13}$, $3 \times 10^{13}$, and $6 \times 10^{13}$ Wcm$^{-2}$ intensities at 4.73 a.u. and 7.0 a.u. nuclear distances and derived the HHG spectra. Some complexity, minima, and oscillatory patterns appeared in the HHG spectra were investigated in this work by the electron localization, electron nonadiabatic dynamics, spatially asymmetric high-order harmonic emission and the Rabi frequency to better understand the origins of these structures in the HHG spectrum. Our simulations showed that the HHG spectra for the $sin^2$-like laser pulses are generated mostly in falling parts of the laser field and as we expect the HHG spectra become complicated. In contrast, the significant part of the HHG of the trapezoidal laser pulses are generated in the constant amplitude part of the laser pulse. Therefore, the complexity is disappeared and the odd harmonic rule does not violate.

We have shown that the oscillatory patterns of the HHG spectra originate from the oscillatory patterns of the $S_g$ and $S_u$ spectra. These oscillatory patterns of the $S_g$ and $S_u$ spectra are due to the slow oscillation patterns in the time-dependent population of the ground and first excited electronic states that in turn are due to a nonadiabatic electronic behavior of the molecule in response to the rapid change of the laser field.

In addition, in this work, we explored how the minima emerge in the high-harmonic spectrum of $H_2^+$. Our results show that the proposed minimum of HHG in the previous reports is related to the oscillatory patterns in $S_g$ and $S_u$ spectra and as mentioned above these oscillatory patterns are due to nonadiabatic electronic behavior and also we can detect other minima in the HHG spectra. Therefore, the appearance of the minima in the HHG spectra are due to a nonadiabatic response of the electronic wavepacket to the rapidly changing laser field.

Our simulation showed that the time-dependent population of the ground and excited electronic states show a fast oscillation corresponding to the Rabi frequency and a slow oscillation pattern corresponding to the variation of the Rabi frequency due to the intensity variation during the rising and falling edge of the laser pulse. Our results show that these slow and Rabi oscillations are appeared for the both $sin^2$-like and trapezoidal laser field. We showed that the variation in the Raby frequency appeared as the slow oscillations are due to the nonadiabatic electron behavior. Therefore, it can be said the mentioned minimum in the reports is related to the oscillations in $S_g$ and $S_u$, and in turn these oscillations are due to slow variations of the Raby frequency and consequently the nonadiabatic electron behavior.

## V. ACKNOWLEDGMENTS

We acknowledge financial support from Iran National Science Foundation: INSF with Grant No. 96015361. We are grateful to Dr. Hamed Ahmadi for his generosity and helpful discussions.

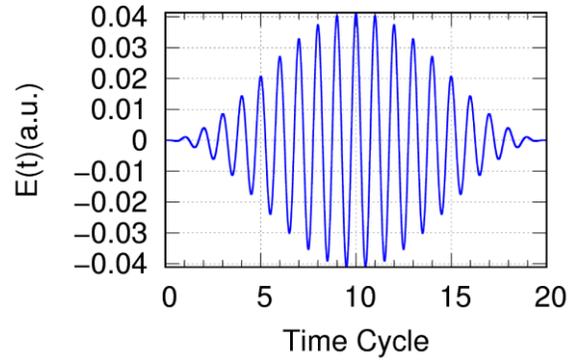

FIG. 1. (Color online) The twenty-cycle sin²-shaped laser pulse with 1400 nm wavelength ($\omega = 0.057$ a.u.) and of $6 \times 10^{13}$ Wcm⁻² intensity.

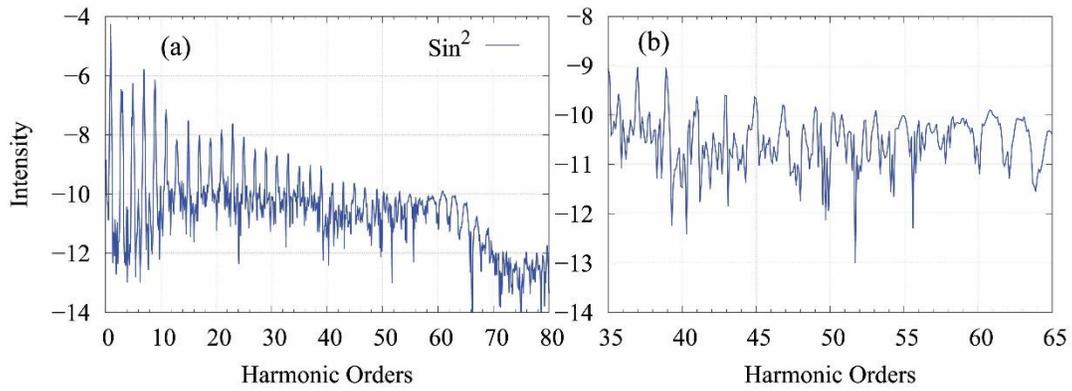

FIG. 2. (Color online) High-order harmonic spectrum for an $H_2^+$ at $R = 7.0$ a.u. internuclear distance under the sin²-shaped laser field with 1400 nm wavelength and $6 \times 10^{13}$ Wcm⁻² intensity (Fig. 1). The right panel shows 35-65 harmonic orders.

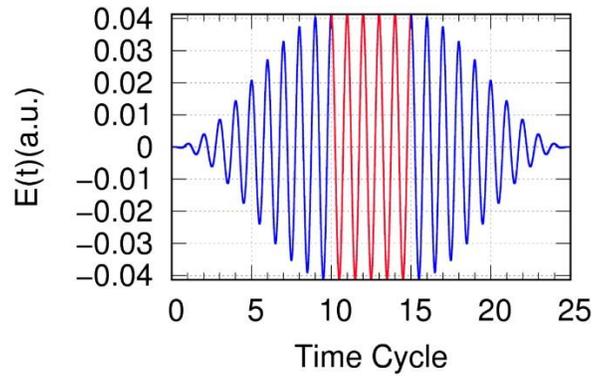

FIG. 3. (Color online) The 25-optical-cycle trapezoidal laser pulse with 10 cycles ramp on, 5 cycles constant, and 10 cycles ramp off (the ramps are sine-squared laser pulse). As compared to Fig. 1, the 5 cycles with constant amplitude (red lines) have been added to the middle of the sin² laser pulse.

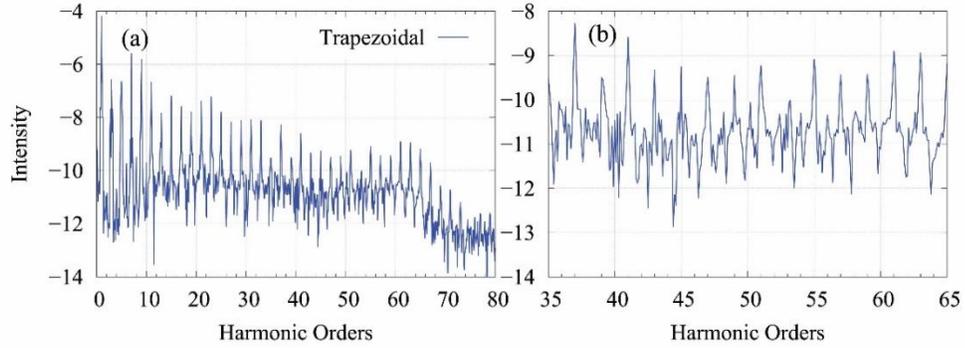

FIG. 4. (Color online) The high-order harmonic spectrum for $H_2^+$ at $R = 7.0$ a.u. internuclear distances, under the trapezoidal laser field shown in Fig. 3, with 1400 nm wavelength and $6 \times 10^{13}$ Wcm$^{-2}$ intensity. The left panel is the total spectrum and the right panel represents the harmonics orders 35-65.

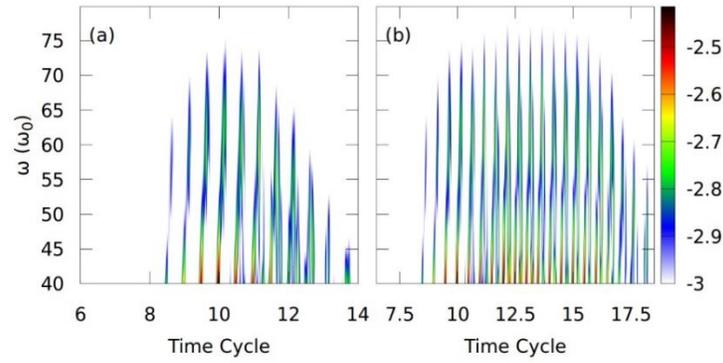

FIG. 5. (Color online) The Morlet-wavelet time profiles for the $H_2^+$ system, in $R = 7.0$ a.u. internuclear distance, under the 20 cycles sin$^2$ laser field (Fig .1) (a) and the 25 cycles trapezoidal laser field (Fig .3) (b) at 1400 nm wavelength and $6\times10^{13}$ Wcm$^{-2}$ intensity. The HHG intensity is depicted in color logarithmic scale on the right side of panels. It can be seen for Fig. 5(a) the harmonics are generated mostly in the falling part of the sin$^2$ laser field (the second ten-cycle) and for the Fig. 5(b), the significant part of the harmonic orders is generated in the constant amplitude of the trapezoidal laser field (10 to 15 cycles).

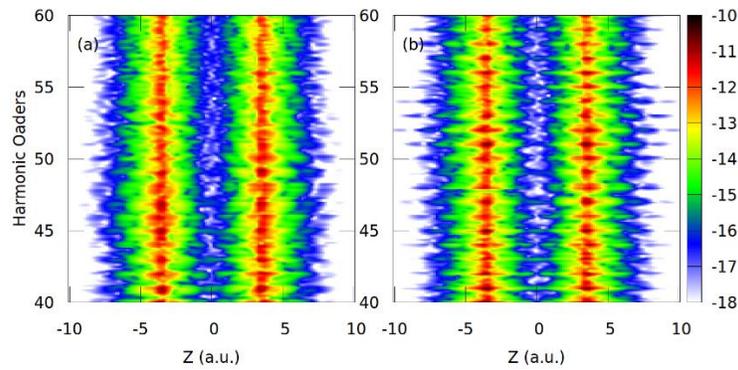

FIG. 6. (Color online) $S(z, \omega)$, spatial distributions of the HHG spectra in terms of the $z$ coordinate and the harmonic orders, of $H_2^+$ with $R = 7.0$ a. u. internuclear distance for the sin$^2$ field (left panel) and trapezoidal field (right panel). The HHG intensity is depicted in color logarithmic scale on the right side of panels.

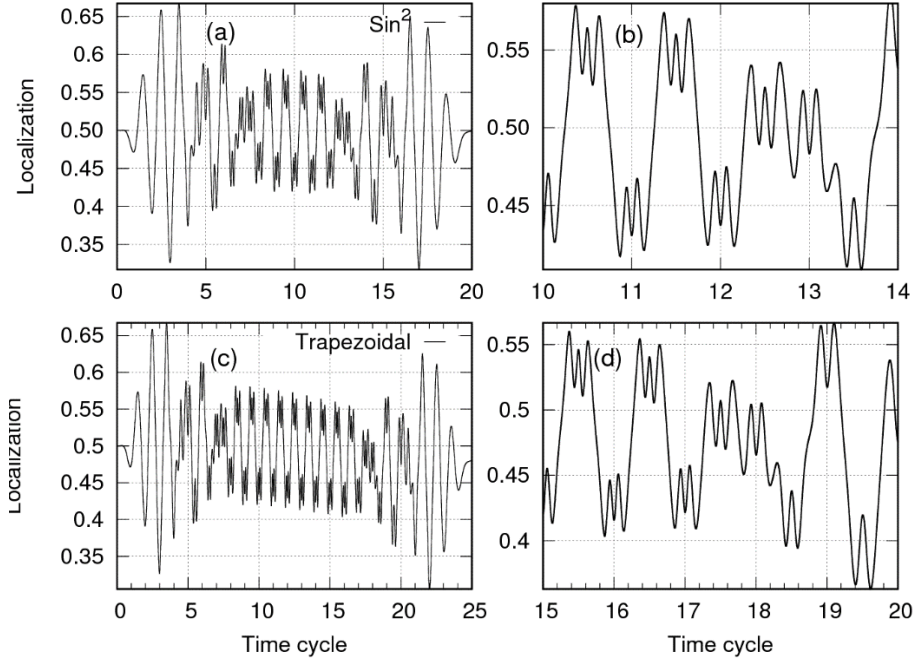

FIG. 7. The electron localization, $\langle z \rangle$, are presented for the sin$^2$-shaped laser field (upper panels) and the trapezoidal laser pulse (bottom panels). For the sin$^2$-shaped laser field (a), adiabatic behavior from 1 to 4 and 16 to 20 cycles and a non-adiabatic behavior from about 4 to 16 cycles. For the trapezoidal field (b), non-adiabatic behavior is from about 4th period until the 20th period, and adiabatic behavior is for others periods (1 to 4 and 21 to 25 cycles). (b) The non-adiabatic behavior of the electron localization maintains a repetitive pattern form about 8th until the 12th periods (near the center of the driving laser), but after this time, this repeating pattern is deformed and for the trapezoidal field(d), the electron localization has non-adiabatic behavior and repetitive trend form about 8th until the 17th periods, but after this time, this repeating pattern is deformed

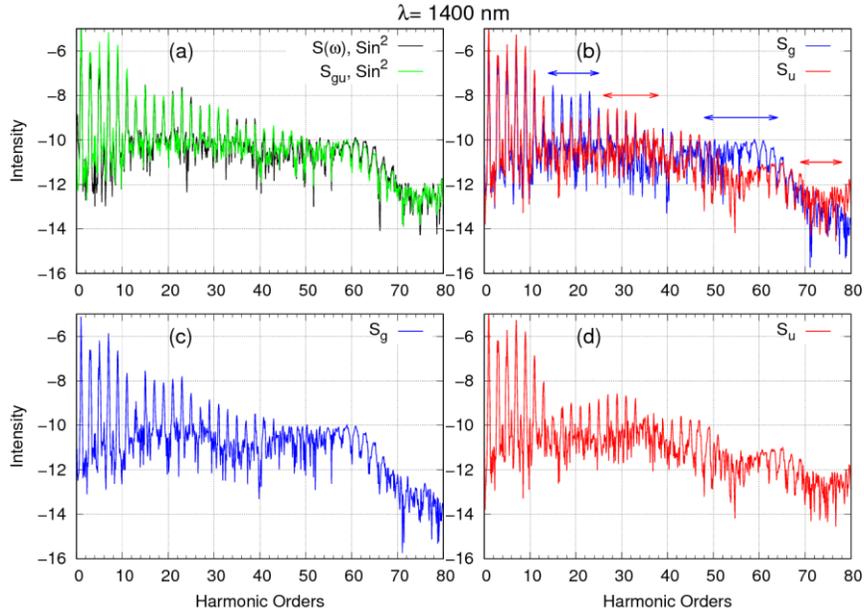

FIG. 8. (a) (Color online) The total harmonic spectrum and (b) the harmonic spectrum both of the ground and first excited electronic states. (c) The harmonic spectrum of the ground and (d) first excited electronic states for the $H_2^+$ in $R = 7.0$ a.u. internuclear distance for the sin$^2$ laser field (Fig. 1) at 1400 nm wavelength.

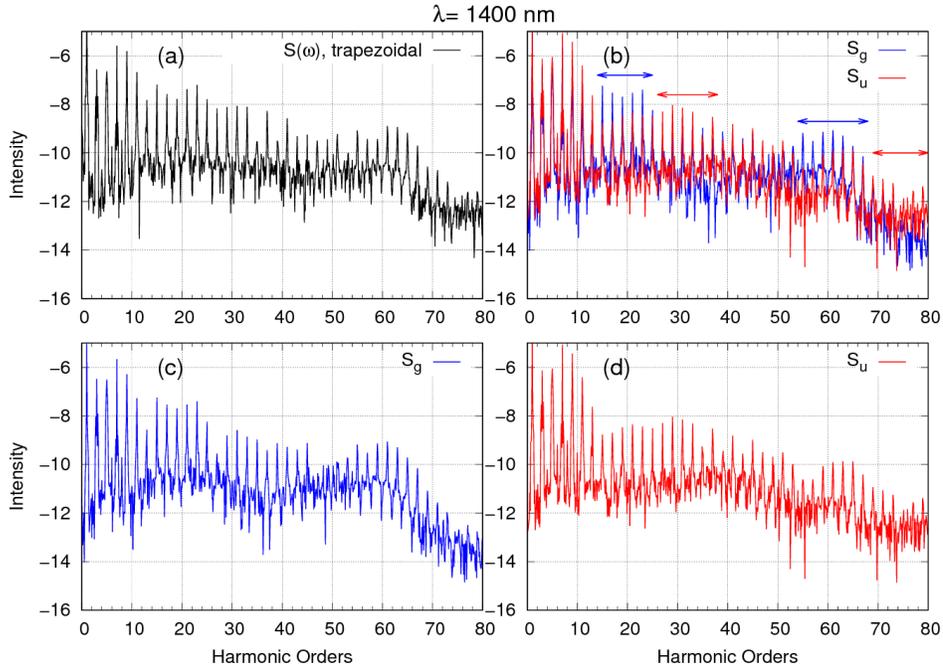

FIG. 9. (Color online) Same as Fig. 8, but for the trapezoidal laser field (Fig. 3).

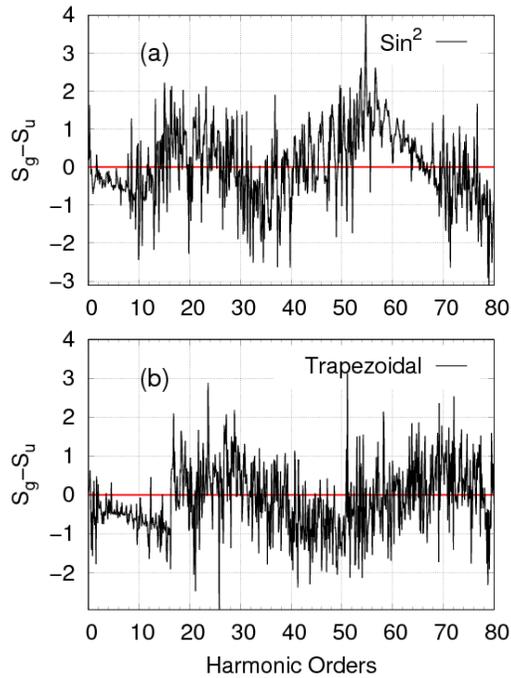

FIG. 10. (Color online) Different between the $S_g$ and $S_u$ ($S_g - S_u$) for $H_2^+$ under the sin$^2$ and trapezoidal laser pulse at 1400 nm wavelength with $I = 6 \times 10^{13}$ Wcm$^{-2}$ intensity.

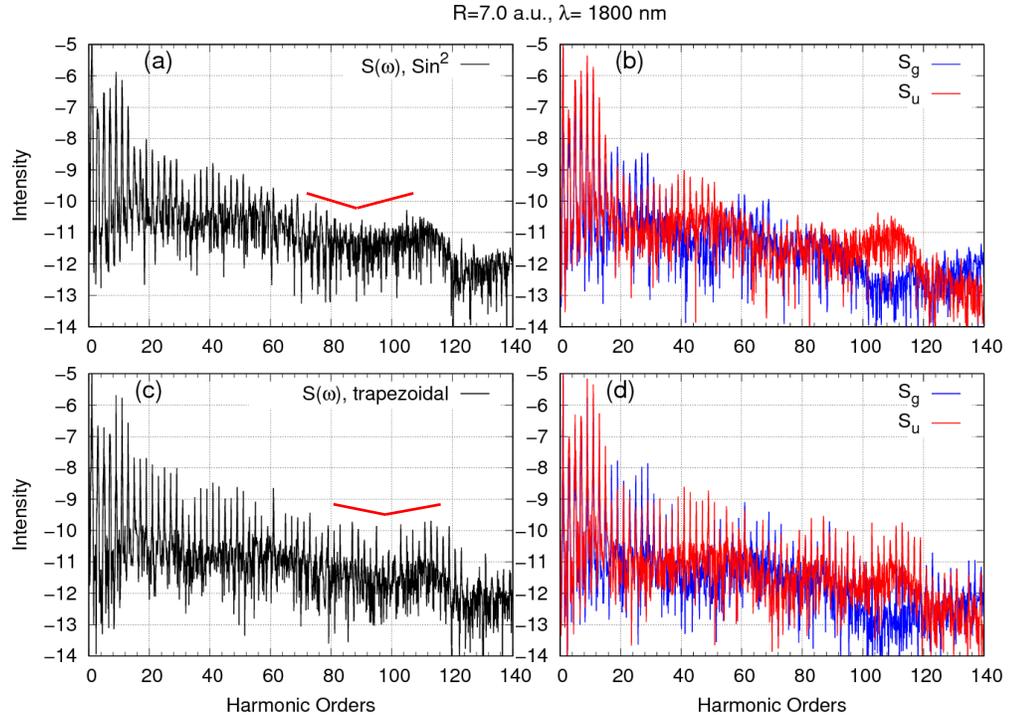

FIG. 11. (Color online) The total HHG spectrum (left panels) and the HHG spectrum due to the ground and first excited electronic states (right panels) for the sin$^2$ and trapezoidal laser pulse for 7.0 a.u. internuclear distance at 1800 nm wavelength with $I = 6 \times 10^{13}$ Wcm$^{-2}$ intensity.

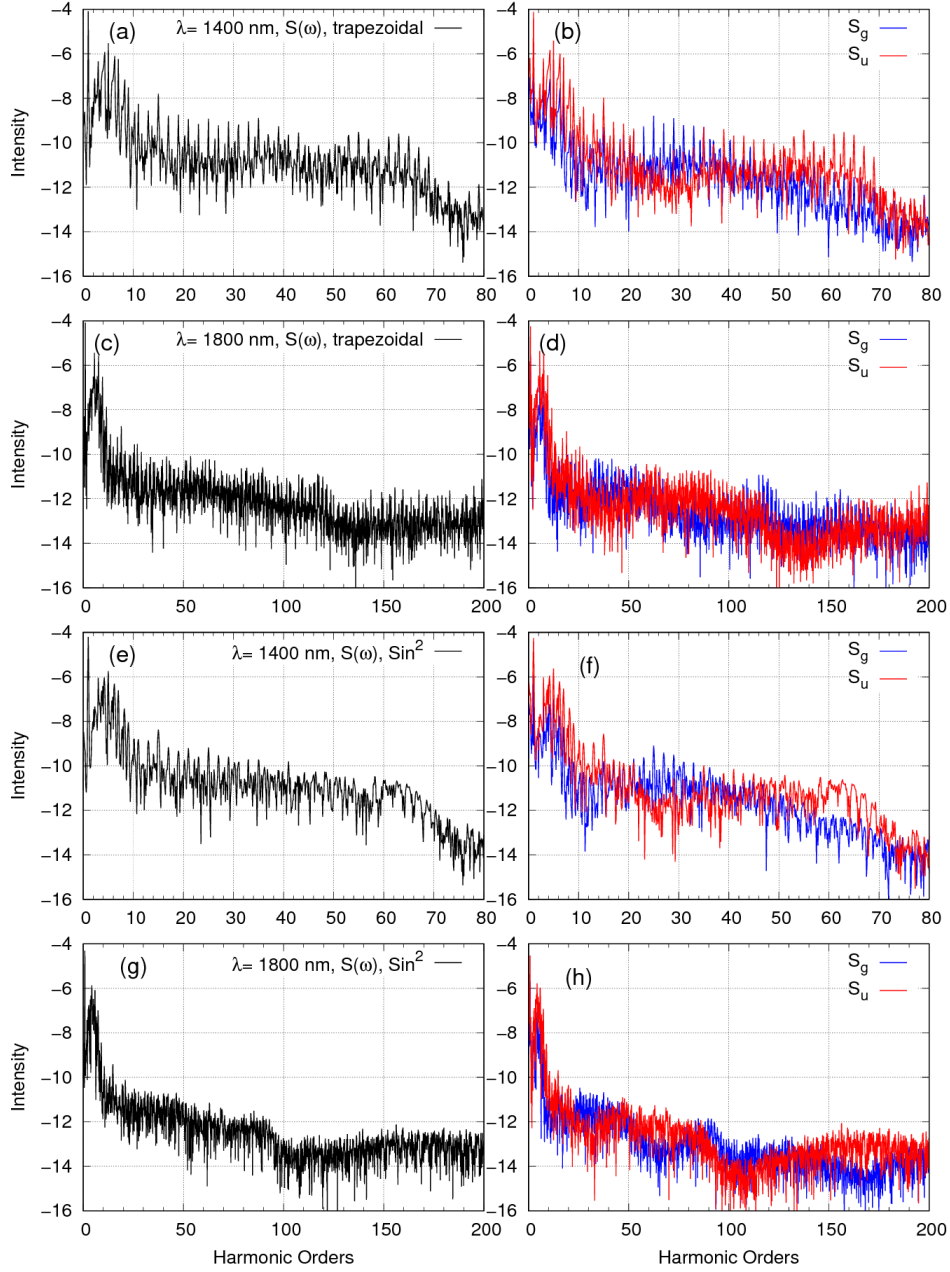

FIG. 12. (Color online) The total HHG spectrum (left panels) and the HHG spectrum of the ground and the first excited electronic states (right panels) for the sin$^2$ and trapezoidal laser pulses with different wavelengths at ~ 4.7 a.u. internuclear distance.

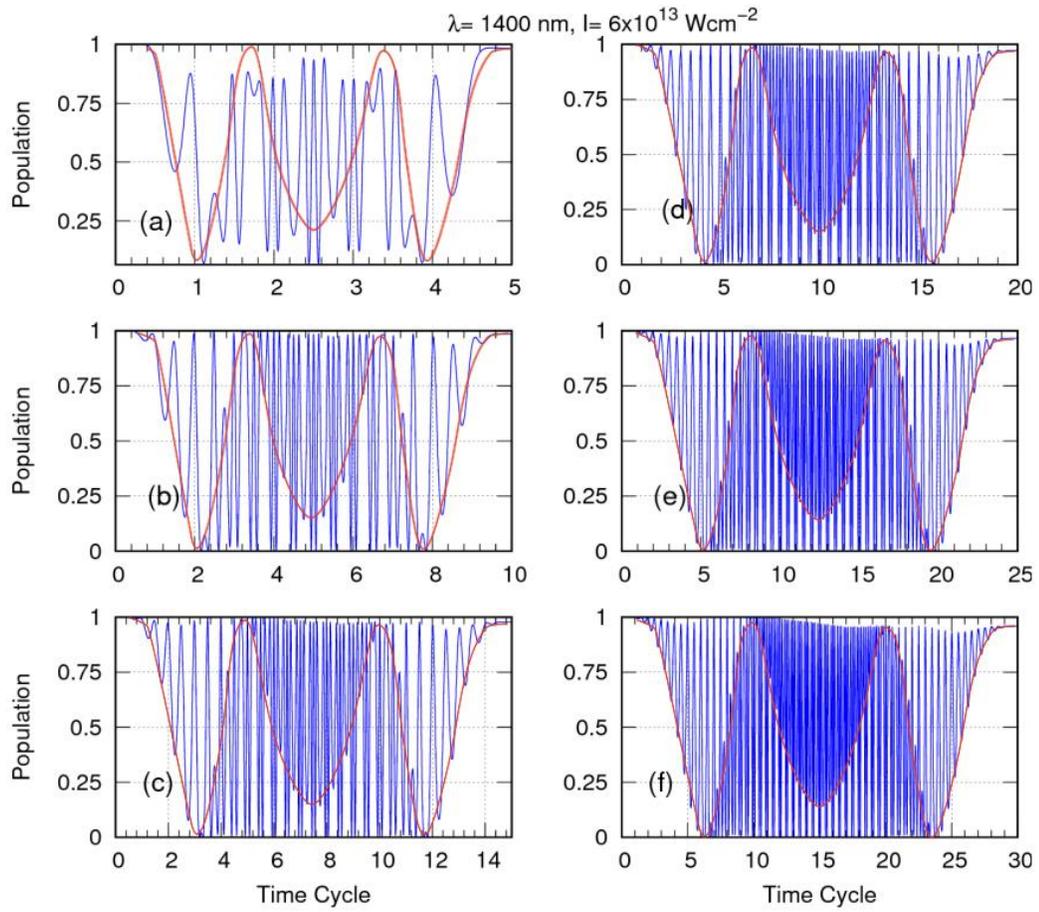

FIG. 13. (Color online) Population of the ground electronic state of $H_2^+$ under the interaction with the laser pulses shown in Fig. 15 with internuclear distance of 7.0 a.u. The populations are shown with blue lines that have fast oscillations with the Rabi frequency. The red lines (the pattern of the slow oscillation in the populations) are related to the Rabi frequency variations.

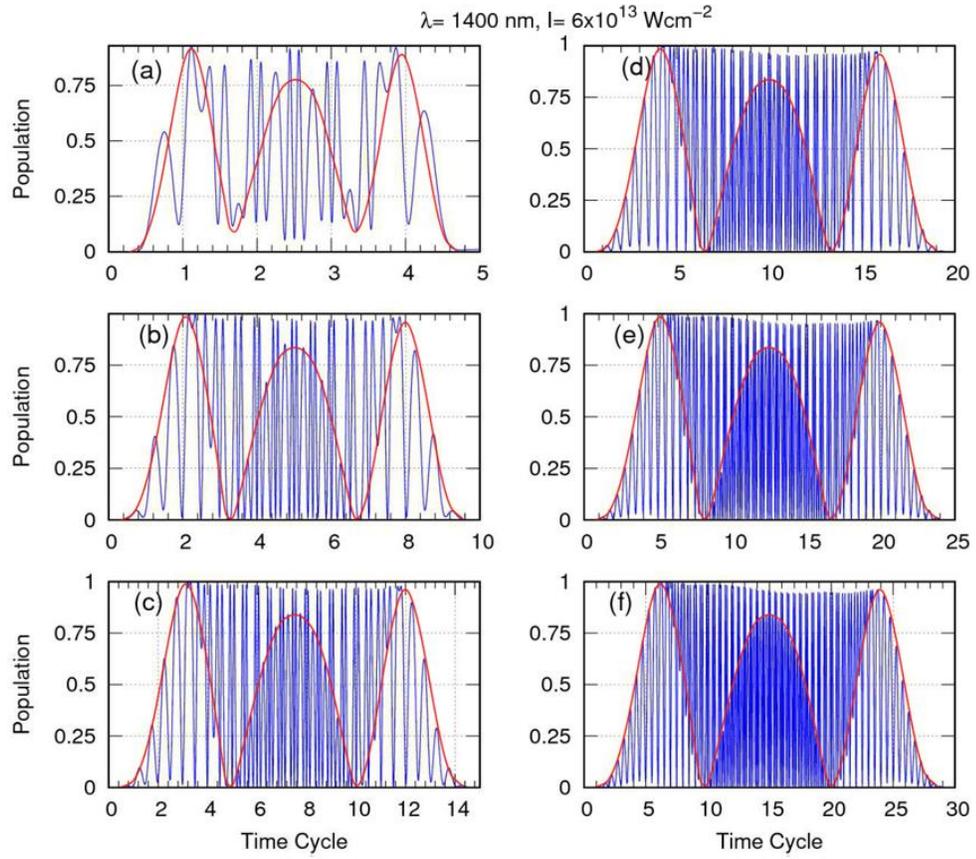

FIG. 14. (Color online) Same as Fig. 13 but for the first excited electronic state.

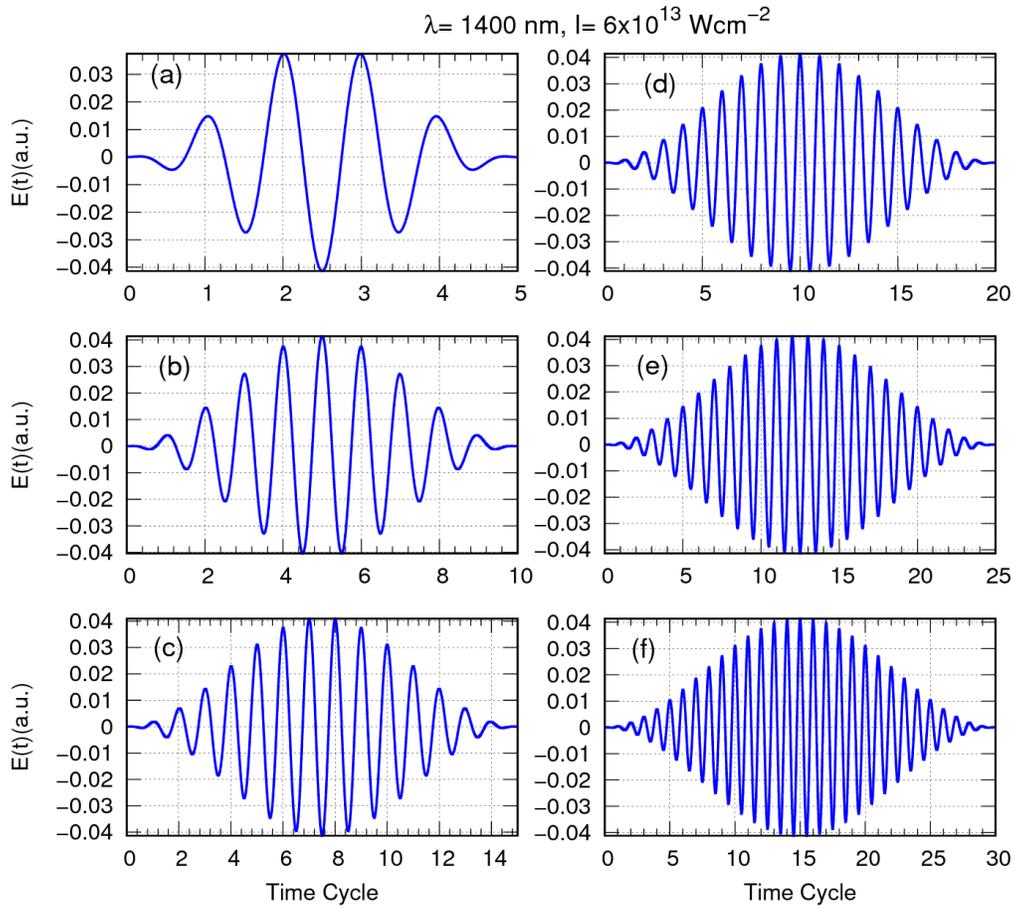

FIG. 15. (Color online) Electric field of $\sin^2$ Laser pulses with different duration at 1400 nm wavelength and $6\times 10^{13}$ Wcm$^{-2}$ intensity used in Fig. 13 and 14.

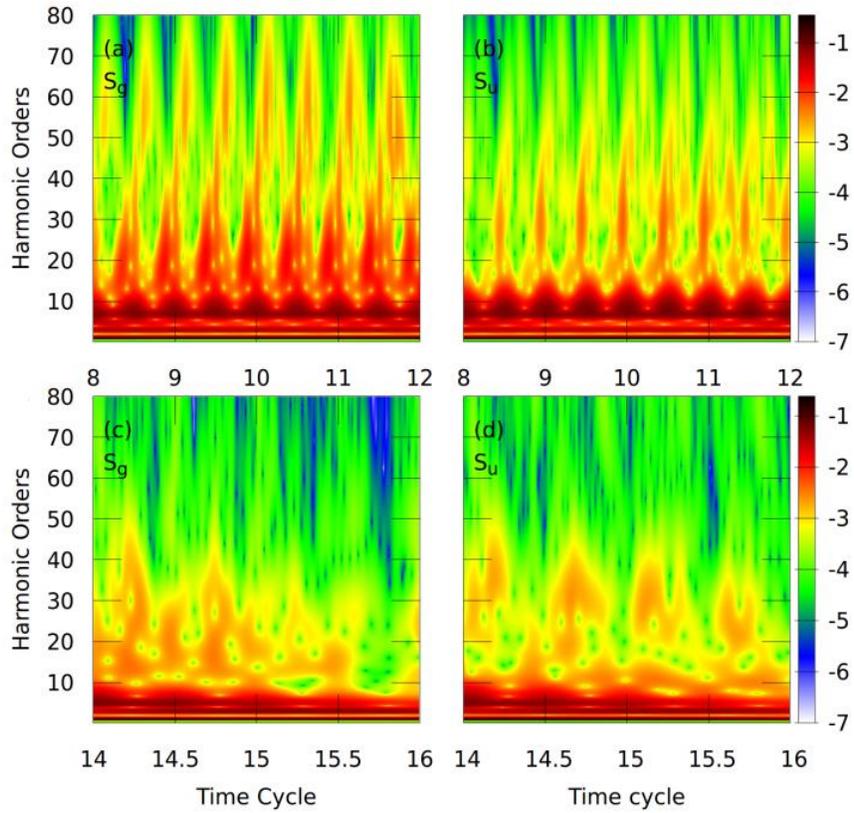

FIG. 16. (Color online) The Morlet-Violet time profiles for 20 cycle $\sin^2$ laser pulses with 1400 nm wavelength and $6\times10^{13}$ Wcm$^{-2}$ intensity for (a,c) the ground electronic state and (b,d) the first excited state at internuclear distances of 7.0 a.u. in different optical cycles. The HHG intensity is depicted in color logarithmic scale on the right side of panels.

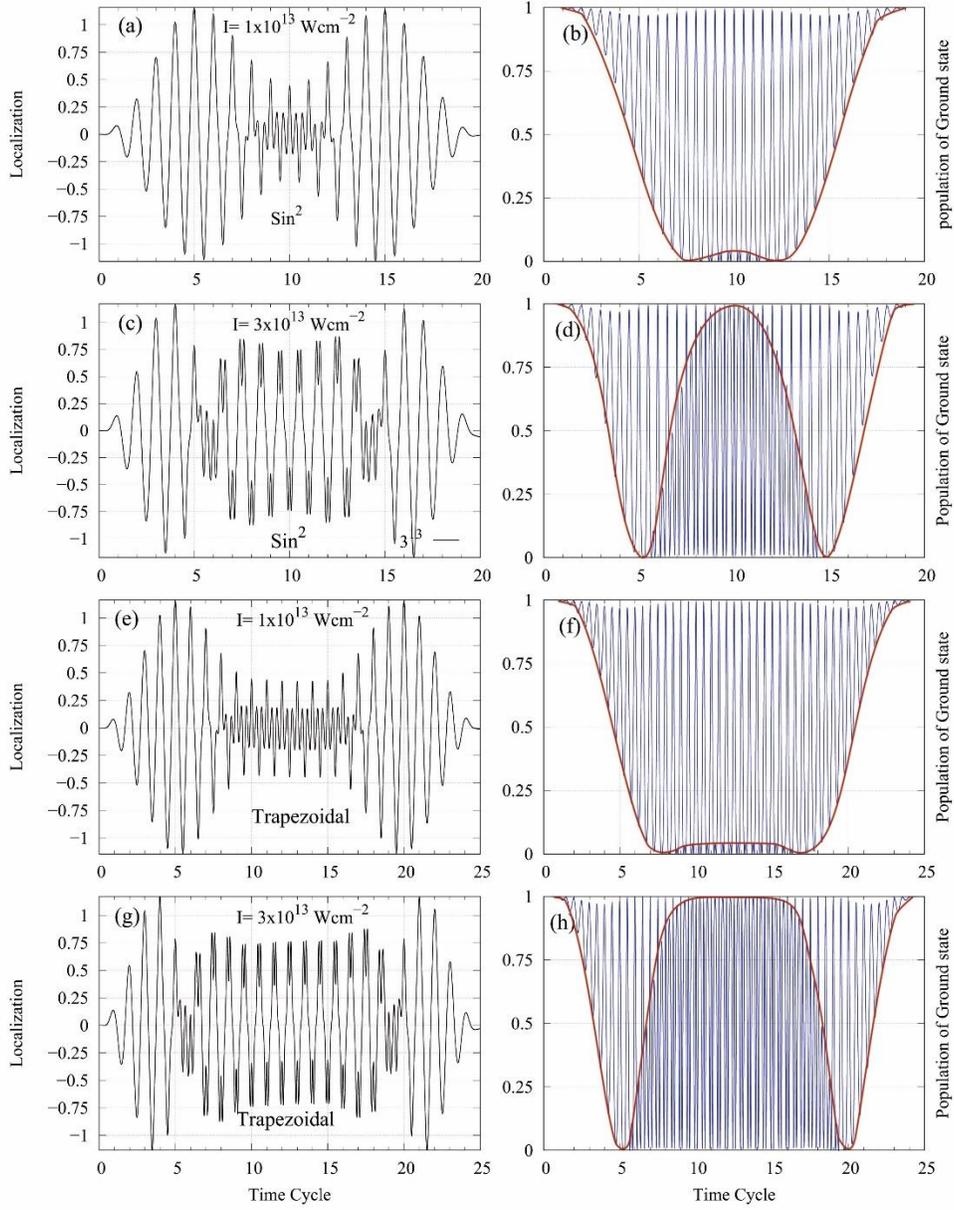

FIG. 17. (Color online) Effect of the intensity of the laser field on the patterns of the electron localization and the ground electronic state population. The electron wave packet localization (a, c, e and g) and the corresponding the time-dependent population of the ground electronic state (b, d, f and h) of the $H_2^+$ system at $R = 7.0$ a. u. internuclear distance under the $\sin^2$-shaped laser pulses and the trapezoidal pulse with the wavelength 1400 nm for different $1\times10^{13}$ and $3\times10^{13}$ Wcm$^{-2}$ intensities. The red lines (the pattern of the slow oscillation in the populations) are related to the Rabi frequency variations.

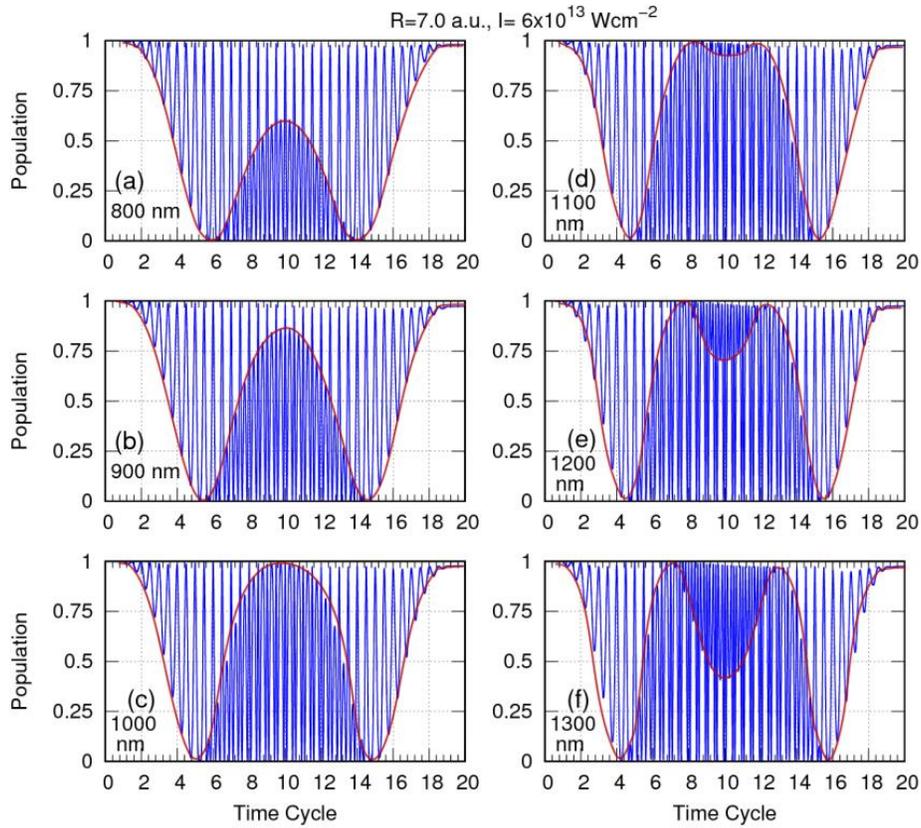

FIG. 18. (Color online) Effect of the magnitude of wavelength on the time-dependent behavior of the ground and excited states population. These figures show the time-dependent population of the ground electronic state of the $H_2^+$ system at $R = 7.0$ a. u. internuclear distance under $\sin^2$-shaped laser pulses at different wavelengths (800, 900, 1000, 1100, 1200, and 1300 nm) and $6\times10^{13}$ Wcm$^{-2}$ intensity. The red lines (the pattern of the slow oscillation in the populations) are related to the Rabi frequency variations.

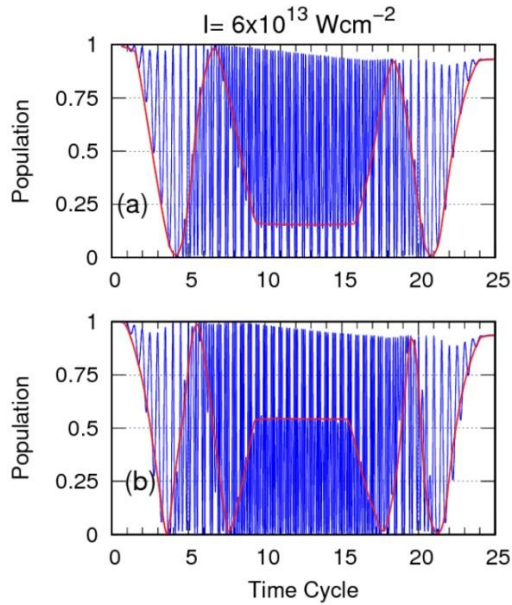

FIG. 19. (Color online) Same as Fig. 18, but for the trapezoidal laser pulses at 1400 nm (a) and 1800 nm (b) wavelengths.

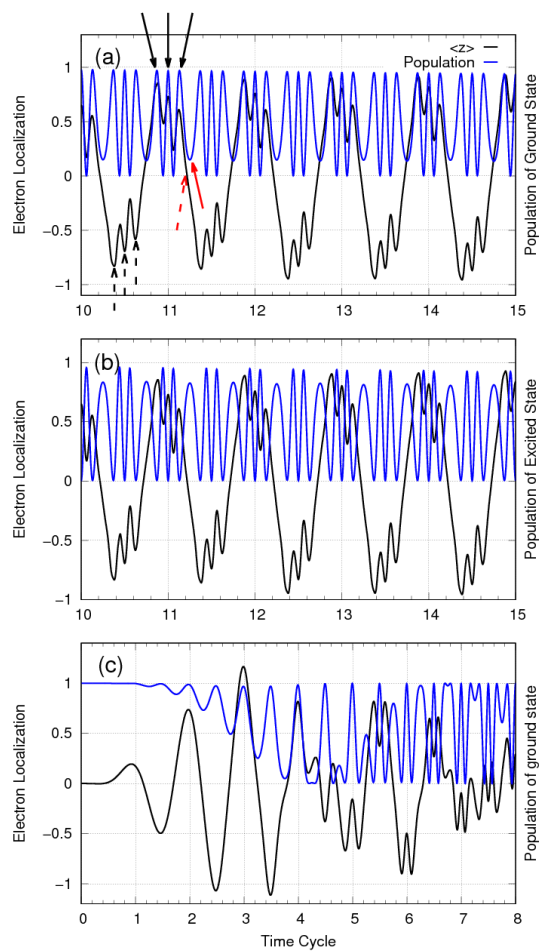

FIG. 20. (Color online) Comparison of the electron localization along the z axis (black line) with the population of the ground and excited electronic states (blue line) for the 20 cycle $\sin^2$ laser pulses at 1400 nm wavelength and $I = 6 \times 10^{13}$ Wcm$^{-2}$ intensity (shown in Fig. 1). It can be seen that when the electron localization is altered from one nucleus (proton) to another nucleus (marked with the red dashed arrow), subsequently a wide valley occurs in the population of the ground electronic state (marked with the red arrow). Also any triplet in the electron localization curve on each nucleus (marked with black dashed arrow) the relate a triplet peak in the population (marked with the black arrow). Similar scenario is also observed for the first excited electronic state shown in panel (b). Panel (c) shows that at the onset of nonadiabatic electronic behavior (at about 4.1 of the time cycle), singlet peaks are replaced by doublet peaks in the electron localization curve.